# ATLAS job monitoring in the Dashboard Framework

J Andreeva[1], S Campana[1], E Karavakis[1], L Kokoszkiewicz[1], P Saiz[1], L Sargsyan[2], J Schovancova[3], D Tuckett[1] on behalf of the ATLAS Collaboration

[1] CERN, European Organization for Nuclear Research, Switzerland

[2] A I Alikhanyan National Scientific Laboratory, Yerevan, Republic of Armenia

[3] Academy of Sciences of the Czech Republic, Prague, Czech Republic

E-mail: Laura.Sargsyan@cern.ch

**Abstract**. Monitoring of the large-scale data processing of the ATLAS experiment includes monitoring of production and user analysis jobs. The Experiment Dashboard provides a common job monitoring solution, which is shared by ATLAS and CMS experiments. This includes an accounting portal as well as real-time monitoring.
Dashboard job monitoring for ATLAS combines information from the PanDA job processing database, Production system database and monitoring information from jobs submitted through GANGA to Workload Management System (WMS) or local batch systems. Usage of Dashboard-based job monitoring applications will decrease load on the PanDA database and overcome scale limitations in PanDA monitoring caused by the short job rotation cycle in the PanDA database. Aggregation of the task/job metrics from different sources provides complete view of job processing activity in ATLAS scope.

## 1. Introduction

The Worldwide Large Hadron Collider Computing Grid [1] provides a large-scale computing environment for the ATLAS experiment. The ATLAS community daily runs more than 600k jobs on the GRID. The number of ATLAS jobs is growing every year and a scalable, robust monitoring system is required in order to follow distributed job processing and to quickly detect and resolve possible problems. Most ATLAS users submit jobs using PanDA (Production and Distributed analysis) [2] workload management system (WMS). A small fraction of jobs are submitted via GANGA [3] job submission system to gLite WMS [4] or local batch system. In order to have a global picture of the ATLAS job processing a system which accumulates job-processing information from different sources into a single place is required. This system also decreases load on the PanDA database, which can then be used solely for operations. The Experiment Dashboard monitoring system [5] that focuses on the needs of the LHC user community was chosen for the implementation since it can meet the challenges of the large-scale ATLAS job processing monitoring. Development work started in the year 2010.

This paper will present an overview of the job monitoring data flow, the functionality of the job monitoring applications, and the future development plans of the new job processing monitoring applications that include task monitoring for production and analysis users and an accounting portal for providing long-term monitoring statistics.

## 2. Job monitoring data flow

2.1. Design overview
ATLAS job processing includes production activity carried out by groups of experts and distributed analysis performed by ATLAS physicists. Dashboard job monitoring uses ORACLE database backend as a data repository. It combines ATLAS job monitoring information from the PanDA job processing database, Production system database, and monitoring information from jobs submitted through GANGA to gLite WMS or local batch systems. GANGA job definitions and jobs running at the Worker Nodes (WN) are instrumented to report job status and meta-information to the Messaging system for the Grid (MSG) [6]. A set of collectors consumes information from different sources and feeds the Dashboard Data Repository. Web User Interfaces at the frontend providing access to monitoring data with different levels of detail are customized to appropriate use cases.

For any access to the database, both read and write, the Experiment Dashboard framework uses the data access object (DAO) pattern as an abstraction layer between server/collector code and the database. Rigorous separation between the database and server/collector code ensures that it is fully agnostic of database implementation, so that any changes to the persistence logic do not impact server/collector code as long as the interface remains correctly implemented.
ATLAS job monitoring data flow is presented in Figure 1.

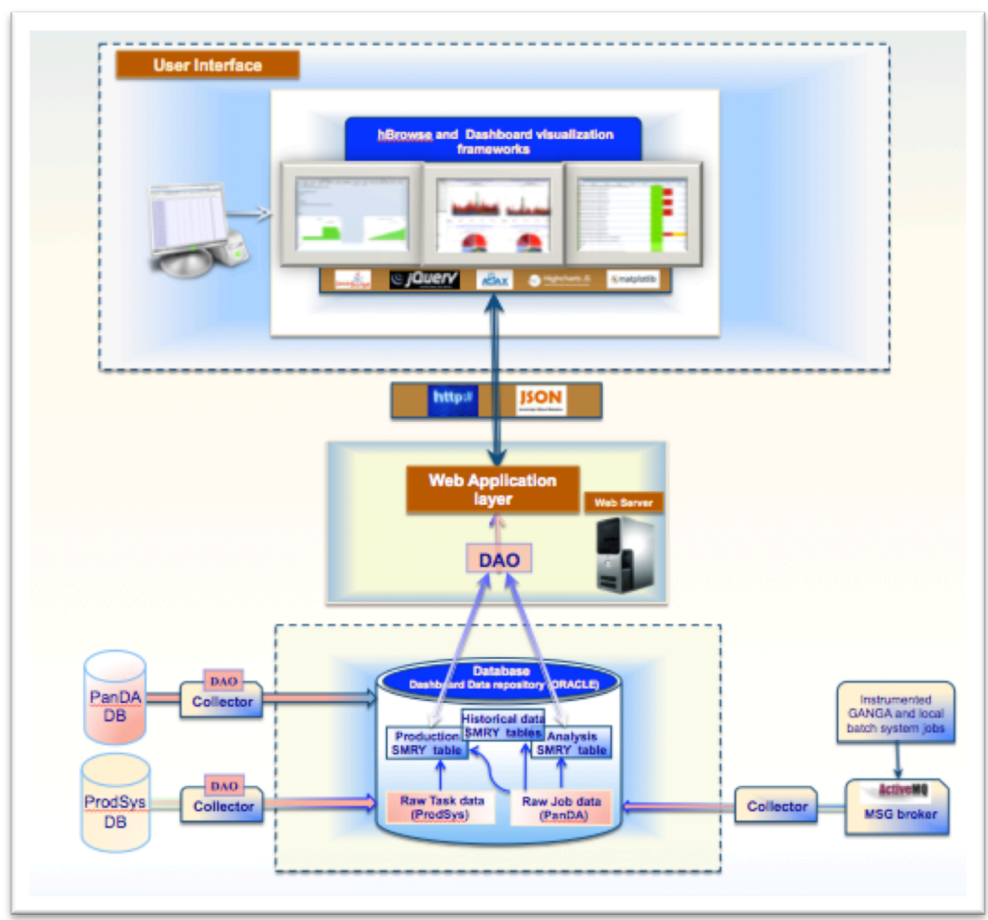

Figure 1. ATLAS job monitoring data flow.

2.2. Main components

2.2.1. *Database*. ATLAS Dashboard database plays the role of a central repository for the production and analysis job monitoring data. As stated above, both read and write access to the DB pass through a data access object (DAO) abstraction layer. Task/job data from consumers is inserted into the Dashboard tables using ORACLE procedures and triggers. Scheduled database procedures aggregate data into summary tables, which are used for high-level overviews. This significantly decreases query response time.

DAO access to the database is done using a connection pool to reduce the overhead in creating new connections, therefore the load on the server is reduced and the performance increased.

2.2.2. *Collectors*. Dashboard collectors query Production and PanDA databases via SQL calls, consume from MSG broker (ActiveMQ) via HTTPS requests, and store job monitoring information in the central Dashboard Data Repository (ORACLE).

In order to ensure the reliability of the service the collectors are being monitored:
- On the database level: DB scheduled jobs. The DB scheduled jobs check the timestamp of the most recent update of the relevant tables and raise an alarm in case of delay.
- On the service collector itself: collector alarms. Collectors check error conditions in the input data flow and raise alarms when errors are detected.
- On the operation system (Linux) level: cronjob scripts. The monitoring scripts check whether collectors are running within foreseen time intervals and restart collectors in case of eventual failure.

In case of detected problems alarms are sent by SMS or e-mail. In case of a collector being stuck or stopped it is restarted.

Consistency checks between PanDA and Dashboard databases, performed on a regular basis, also ensure the reliability of data exposed through the Dashboard UIs.

2.2.3. *Web Application layer*. The data source is fully decoupled from the user interface. Dashboard framework web Application layer provides HTTP entry point to the data and serves information in XML (extensible markup language), CSV (comma separated values), JSON (JavaScript object notation) or image formats. Third-party clients can also consume these serialized data.

2.2.4. *User interface*. Depending on the application the user interface is implemented using either Experiment Dashboard or hBrowse [7] visualization frameworks.

The hBrowse based user interface for the Analysis and Production Task Monitor is a client-side JavaScript application that communicates with the server using asynchronous JavaScript requests from the client browser (AJAX) and expects JSON responses. It utilizes modern web technologies such as jQuery JavaScript library [8], HighCharts plotting library, DataTable and SearchTable jQuery plugins, and Google charts library.

## 3. Job monitoring applications

3.1. Analysis Task Monitor

The Experiment Dashboard Analysis Task Monitoring application [9] is a user-centric monitoring tool. It collects and exposes information, which describes the progress of the processing of user tasks. Typically, analysis calculations are performed in large chunks by jobs which have a similar setup, i. e.

they perform specific processing over a given data collection. Such a group of jobs is called an analysis task. Each job of a given task processes some portion of an input data collection and produces a set of output files. According to ATLAS definition, a complete set of output files produced as a result of specific processing over a given data collection, represents some output container.

User interface (UI) [10] provides access to the user's tasks, using a secure web connection (https) and Grid Certificate. However a demo version allows users to try out the functionality without being authenticated.

Functionality:
- Monitoring on the task (collection of jobs, based on output container) level and individual job level.

User interface includes sorting and filtering possibilities:
- Tasks can be filtered by name or some pattern inside the name or by time ranges.
- Graphical representation:
    o Job Evolution, jobs distribution by site, distribution of failed jobs by failure error category (Figure 2).
- Summary charts for task and job tables.

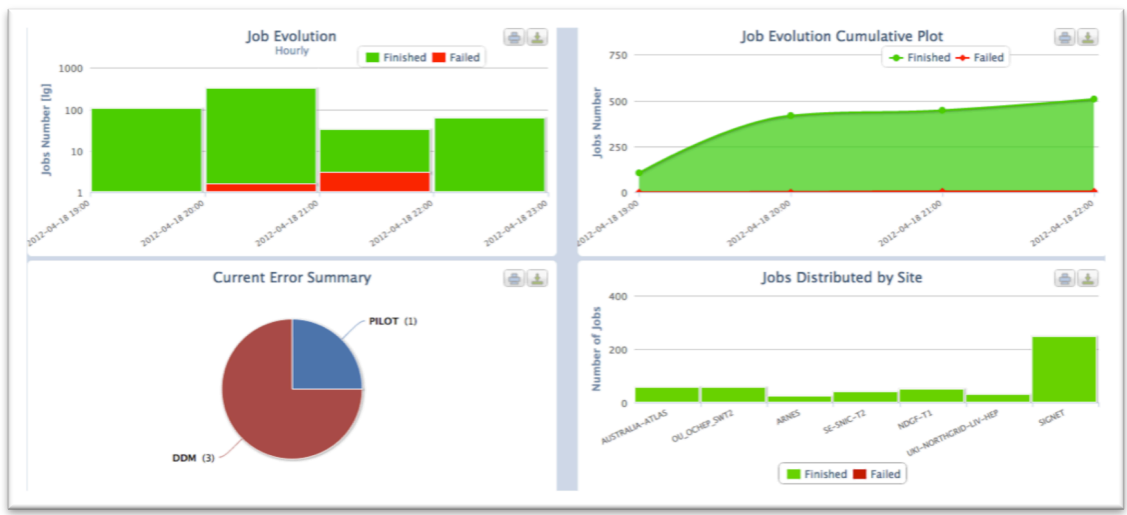

Figure 2. Analysis Task Monitor: graphical plots under a task expanded row.

The initial page presents a list of tasks that the user has submitted in the last day. Task meta information (creation time of the task, last modification time, input dataset used, list of sites where the jobs of this particular task are running, build job(s) number(s), etc.) (Figure 3).

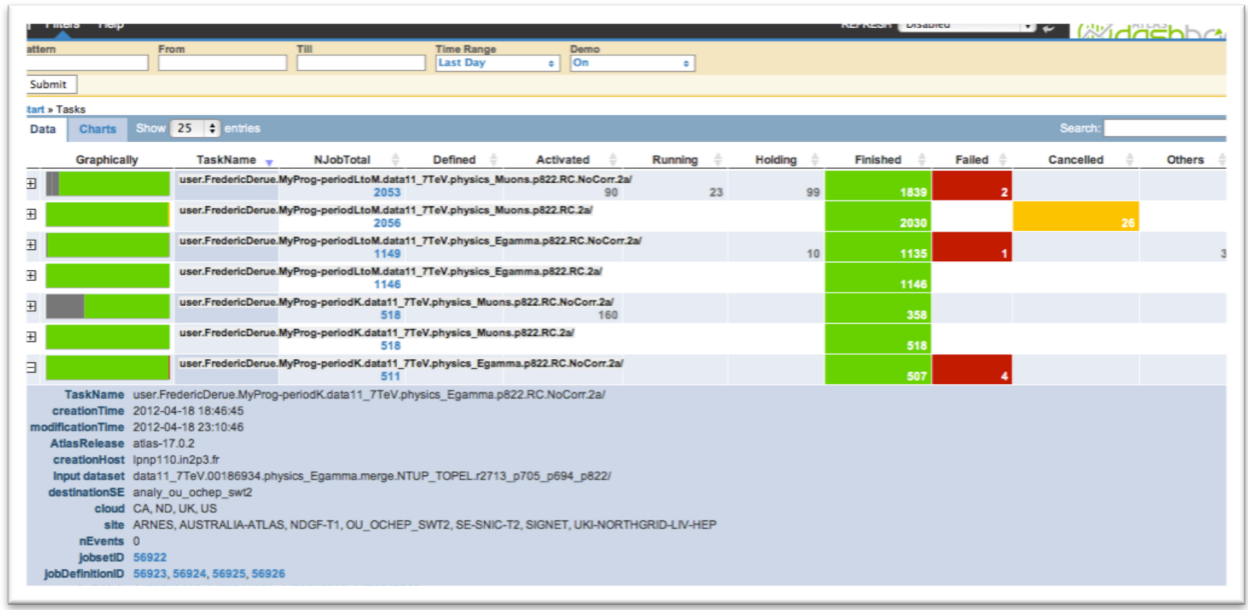

Figure 3. The user interface of the Analysis Task Monitor: filters, task table, and task meta information.

Advanced interactive plots could be exported as an image or PDF document.

The Task Monitoring information includes the job status of individual jobs in the task, their distribution by site and over time, the reason of failure, the number of processed events and the resubmission history.

The application is used by analysis users and by the analysis support team.

3.2. Production Task Monitor

The Production Task Monitor application aims to satisfy the needs of the production operators, production managers and site administrators. Production UI [11] allows users to follow the progress of their production tasks by offering a wide selection of statistics and graphical plots (Figure 4).

By using this tool, production operators can easily identify the problematic tasks, which require action, or detect sites with incorrect configuration, for example with incorrect software installation. The application provides the possibility to get the categorized error statistics for each task.

Production managers and site administrators can benefit from functionality like a multifunctional filter that provides a co-dependency capability for enabled filters (Figure 5). This means that selection of a particular value in one filter automatically disables invalid values in another filter, in case of any dependency between them. Moreover, the privileged users can predefine set of filter values (time range, working group, cloud, pattern, etc.) to create group specific requests by editing a twiki page [12]. Privileged users can also make changes in Help topics, using twiki page [13].

The application is being developed in close collaboration with the ATLAS production managers.

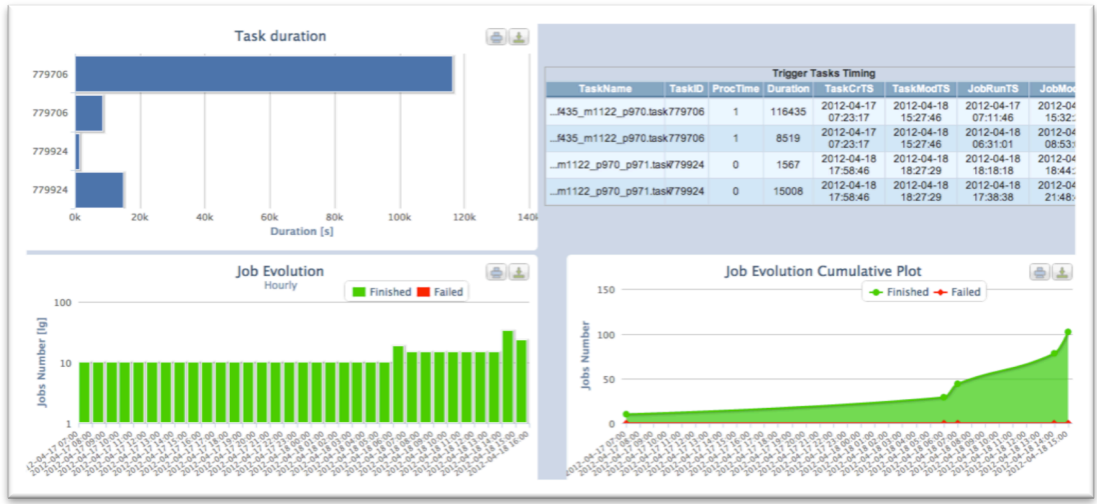

Figure 4. Production Task Monitor: graphical plots and timing table.

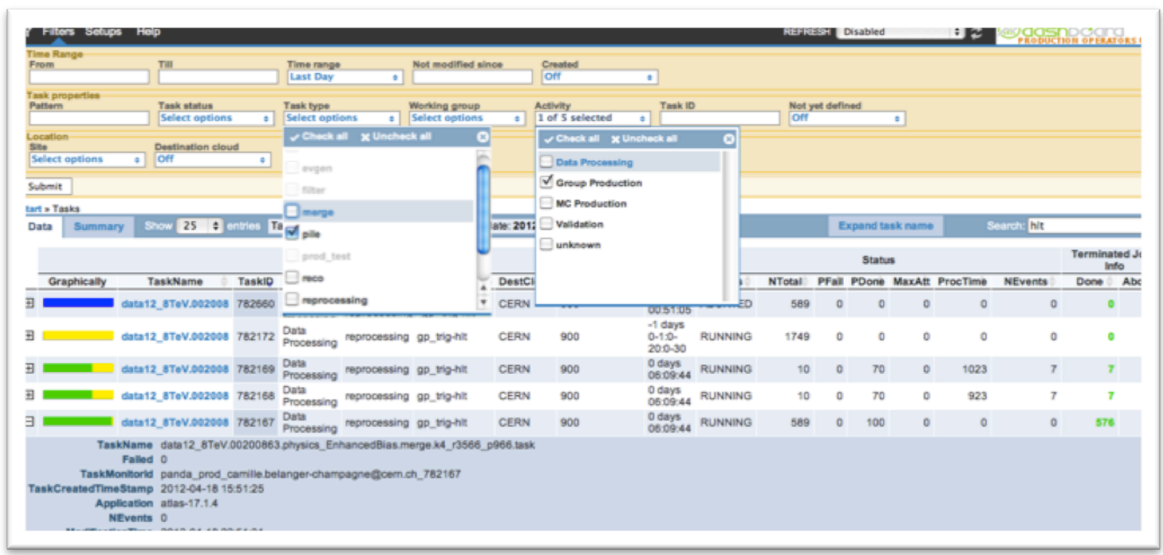

Figure 5. The user interface of the Production Task Monitor.

3.3. Historical Views Monitor

The Historical Views application is an accounting-style UI [14] that presents job processing monitoring metrics and resource utilization information as a function of time. The interface uses aggregated tables in the database to provide efficient access to historical information.
   Functionality:
- Number of completed, submitted (Figure 6), pending, running jobs.
- Number of successful, failed jobs, efficiencies based on successful/accomplished jobs.
- Distribution of failed jobs by failure code, reason, category.

- CPU/Wall clock consumption, efficiency as CPU versus wall clock time.
- Number of processed events as a function of time, CPU/wall clock time spent on a single event.
- Resource utilization, number of used slots, efficiency of site usage compared to pledges.
- Activities at the site. Single site view with job processing metrics.
- All data can be filtered by site, tier, country, activity, processing and destination cloud, data type, project and group.
- Any time range can be selected.
- Available granularities are hourly/daily/weekly/monthly.
- All data is available in machine-readable format (xml, csv, json).
- All plots and distributions in machine-readable format are available via direct link.

The web UI provides a wide selection of graphical plots for selected parameters and serves to understanding the nature of the infrastructure inefficiencies, reasons of failures, and helps to resolve and predict problems.

The Historical Views application is widely used by site admins, and VO computing managers.

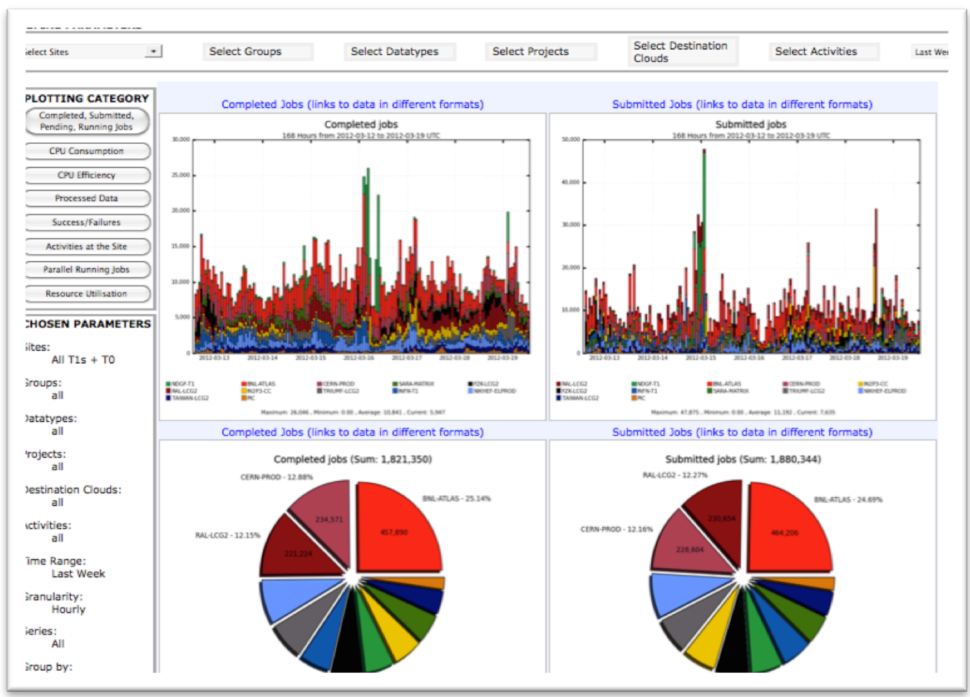

Figure 6. Historical Views: terminated, submitted jobs.

## 4. Future plans

One of the important advanced functionality which is foreseen by the development plans is enabling of cancellation and resubmission on the task and individual job level from the Analysis Task Monitoring web user interface. For example through Analysis Task Monitor a user will be able to not only identify that a job have failed but will also be able to interactively kill task/jobs and resubmit them to a different site if a site experiences problems.

ATLAS job processing monitoring applications will be extended according to the needs of the ATLAS user community.

## 5. Conclusions

ATLAS Job Processing monitoring applications provide a complete view of job processing activity in the scope of ATLAS. Regular consistency checks between PANDA DB and Dashboard ensure data reliability. Advanced and intuitive user interfaces, providing a wide selection of graphical plots, filtering possibilities and good performance, have proved to satisfy the needs of various categories of ATLAS users: site admins, ATLAS VO managers, analysis users, user support teams, production operators and managers, and shifters [15]. Dashboard job monitoring applications have become an essential component of ATLAS distributed computing operations.